\RequirePackage{amsmath}
\documentclass[12pt]{iopart}
\pdfoutput=1 %to include pdf pictures/figures
\usepackage{graphicx,cite}
\usepackage{amsmath,amsfonts,amssymb}
\usepackage{bm,bbm}
\usepackage[urlcolor=blue]{hyperref}
\usepackage{graphicx}
\usepackage[T1]{fontenc}

\newcommand{\f}[1]{\mathbf{#1}}
\newcommand{\Elr}[1]{\left\langle #1\right\rangle}
\newcommand{\E}[1]{\langle #1\rangle}
\newcommand{\Es}[1]{\left\langle #1\right\rangle_{\rm s}}
\newcommand{\ps}{p_{\rm s}}
\newcommand{\js}{\mathbf{j}_{\rm s}}

\newcommand{\var}{{\rm var}}
\newcommand{\cov}{{\rm cov}}
\newcommand{\vars}{{\rm var_{s}}}
\newcommand{\covs}{{\rm cov_{s}}}
\newcommand{\sigmavar}{{\rm var}_{\Sigma}}
\newcommand{\sigmaE}[1]{\left\langle #1\right\rangle_{\Sigma}}
\newcommand{\x}{\f x}

\newcommand{\abs}[1]{\left\lvert #1 \right\rvert}

\newcommand{\bsig}{\boldsymbol{\sigma}}

\newcommand{\bnu}{{\boldsymbol{\nu}}}
\newcommand{\bnus}{{\boldsymbol{\nu}_{\rm s}}}

\renewcommand{\P}[1]{\mathbb{P}\left[#1\right]}
\newcommand{\R}{\mathbb{R}}
\newcommand{\N}{\mathbb{N}}
\newcommand{\C}{\mathbb{C}}

\newcommand{\pvar}{{\rm pvar}}

\renewcommand{\x}{x}

\renewcommand{\js}{j_{\rm s}}
\renewcommand{\bnu}{\nu}
\newcommand{\nus}{\nu_{\rm s}}
\renewcommand{\bnus}{\nu_{\rm s}}
\renewcommand{\f}[1]{#1}
\usepackage{xcolor}
\definecolor{C0}{rgb}{0.12156862745098039, 0.4666666666666667, 0.7058823529411765}
\definecolor{C1}{rgb}{1.0, 0.4980392156862745, 0.054901960784313725}
\definecolor{C2}{rgb}{0.17254901960784313, 0.6274509803921569, 0.17254901960784313}
\definecolor{C3}{rgb}{0.8392156862745098, 0.15294117647058825, 0.1568627450980392}
\definecolor{C4}{rgb}{0.5803921568627451, 0.403921568627451, 0.7411764705882353}
\definecolor{C5}{rgb}{0.5490196078431373, 0.33725490196078434, 0.29411764705882354}
\definecolor{inferno0}{rgb}{0.087411, 0.044556, 0.224813}
\definecolor{inferno1}{rgb}{0.379001, 0.076253, 0.432719}
\definecolor{inferno2}{rgb}{0.658463, 0.178962, 0.372748}
\definecolor{inferno3}{rgb}{0.894305, 0.353399, 0.193584}
\definecolor{inferno4}{rgb}{0.987622, 0.64532 , 0.039886}

\colorlet{mylinkcolor}{blue!66!black!80}
\hypersetup{colorlinks = true, linkcolor = {mylinkcolor}, linkcolor = {mylinkcolor}, citecolor = {mylinkcolor}, urlcolor = {mylinkcolor}}
\definecolor{grey}{rgb}{0.6,0.6,.6}
\definecolor{darkgrey}{rgb}{0.4,0.4,.4}
\definecolor{darkgreen}{rgb}{0,0.4,0}
\definecolor{lightgreen}{rgb}{0,0.7,0}
\definecolor{darkred}{rgb}{0.5,0,0}

\newcommand{\blue}[1]{{\color{blue}#1}}

\renewcommand{\blue}[1]{{#1}}

\allowdisplaybreaks

\begin{document}
\title[Finite-time thermodynamic correlation inequalities]{Thermodynamic correlation inequalities for finite times and transients}
\author{Cai Dieball and Alja\v{z} Godec}
\address{Mathematical bioPhysics Group, Max Planck Institute for Multidisciplinary Sciences, 37077 G\"ottingen, Germany}
\ead{agodec@mpinat.mpg.de}

\begin{abstract}
Recently, a thermodynamic bound on correlation times was formulated in
[A. Dechant,  J. Garnier-Brun, S.-i.\ Sasa, Phys.\ Rev.\ Lett.\ 131,
  167101 (2023)], showing how the decay of correlations in Langevin
dynamics is bounded by short-time fluctuations and dissipation. 
Whereas these original results only address very long observation
times in steady-state dynamics, we here generalize the respective
inequalities to finite observations and general initial conditions. We utilize
the connection between correlations and the fluctuations of
time-integrated density functionals and generalize the direct
stochastic calculus approach from [C. Dieball and A. Godec,
  Phys. Rev. Lett. 130, 087101 (2023)] which paves the way for further
generalizations. We address the connection between short and long time
scales, as well as the saturation of the bounds via complementary
spectral-theoretic arguments. Motivated by the spectral insight, we
formulate all results also for complex-valued observables.
\end{abstract}

\section{Introduction}
Correlations are a central property of stochastic systems with
non-trivial time evolution \cite{vanKampen2007}. The temporal extent of
correlations is generally considered to be a system-specific quantity. However, 
in the particular
case of overdamped diffusions, correlations of observables---formally
functions on the respective state space---have recently been found to be bounded \emph{generically} by the purely diffusive
short-time fluctuations under the assumption of strongly ergodic long-time
$t\to\infty$ dynamics \cite{Dechant2023PRL} (conceptually related results were also
obtained for Markov-jump dynamics \cite{Hasegawa2024PRL}). These works
connect the decay of correlations to stochastic thermodynamics.
Moreover, it is known that correlations in non-equilibrium steady
states (NESS) may relax faster than in equilibrium systems (see, e.g.,
\cite{ReyBellet2015,Kapfer2017PRL} where the authors \blue{mainly used}
large-deviation ($t\to\infty$) ideas \cite{ReyBellet2015,Coghi2021PRE}). 

More broadly, the thermodynamic entropy production is central to
stochastic thermodynamics \cite{Seifert2012RPP,Dieball2025JCP}. Inference of the
entropy production has been addressed in the context of thermodynamic
uncertainty relations
(TURs)~\cite{Barato2015PRL,Gingrich2016PRL,Horowitz2019NP,Liu2020PRL,Hartich2021PRL,Koyuk2022PRL}
(that were further improved via spatial coarse graining
\cite{Dieball2022PRL,Dieball2022PRR}, by incorporating correlations
\cite{Dechant2021PRX,Dieball2023PRL}, or focusing on the short-time
limit \cite{Manikandan2020PRL}), as well as 
from splitting probabilities for transitioning between 
\cite{Hartich2021PRX,Blom2024PNASU,Tassilo}, and waiting-time statistics within~\cite{Skinner2021PRL,Skinner2021PNASU,vanderMeer2022PRX,Harunari2022PRX}, coarse mesostates, 
from trajectory snippets \cite{vanderMeer2023PRL,Deguenther2024PNASU}, via neural networks \cite{Kim2020PRL,Otsubo2022CP}, 
and from speed and transport limits \cite{YoshimuraPRL2021,VanVu2023PRX,Leighton2022PRL,Leighton2023PRL,Cai_transport,Shiraishi2018PRL,CSL_4}, 
or the variance sum rule \cite{DiTerlizzi2024S}. \blue{Further recent work demonstrated the utility of a deep learning framework to infer entropy production \cite{Boffi2024PNASU} and connected intricate spectral properties of Langevin dynamics with linear drift to entropy production \cite{Fyodorov2025PRL}.}

A promising complementary approach is the inference of entropy
production from the decay of correlation functions of observables,
which has so far been demonstrated for the large-time limit
$t\to\infty$ in Ref.~\cite{Dechant2023PRL}. This last approach is
promising in particular because it even allows to infer entropy
production in cases where the irreversible driving is fully hidden and
the observed dynamics (i.e., a projection of the full system's dynamics)
is perfectly time-reversal symmetric (see example in
Ref.~\cite{Dechant2023PRL}). However, the results in
\cite{Dechant2023PRL} are not valid for finite (i.e., short) times nor
for transients, which somewhat limits their application.

To fill this gap we here generalize the results of
Ref.~\cite{Dechant2023PRL} to finite times and transient dynamics
settling into equilibrium or detailed balance (DB) and non-equilibrium
steady states (NESS), respectively. Moreover, we extend the results to
complex-valued observables to tighten the bounds used for inference.  

\section{Summary of the results in Ref.~\cite{Dechant2023PRL}}
First,  we summarize the main results from Ref.~\cite{Dechant2023PRL}
which we then generalize below. We adapt the notation in order to optimize
it for the general setting and to make the proof as easy as possible.

Consider scalar observables $z_t=z(\x_t)$ and $\chi_t=\chi(\x_t)$ which are
square integrable functions $z, \, \chi:\Omega\to\R$ of the full $d$-dimensional
time-homogeneous overdamped Markovian dynamics 
\begin{align}
    \rmd\x_\tau = a(\x_t)\rmd\tau+\sigma\rmd W_\tau \label{SDE}
\end{align}
in a domain $\Omega\subseteq \R^d$. Here, $\rmd W_\tau$
denotes the increments of the Wiener process, $D=\sigma\sigma^T/2$ is
positive definite, and $a:\Omega\to\R^d$ is the drift that will later be restricted to give rise to DB or NESS dynamics.

In Ref.~\cite{Dechant2023PRL} the following results were shown for DB
or NESS dynamics, respectively, where the initial condition is sampled
from the steady state.  Note that all results involve time-integrals
from $t=0$ to $t=\infty$, i.e., in principle they only apply to
infinitely long trajectories (in practice, trajectories that are much longer
than the relaxation time).   
Averages are taken with respect to the steady-state measure and are
denoted by $\Es{\cdots}$. The covariance is defined as
$\covs(z_t,z_0)\equiv\Es{z_t z_0}-\Es{z_t}\Es{z_0}$. Note that
$\Es{z_t}=\Es{z_0}$ for all $t\ge 0$ and hence we may use simplified
notations like $\Es{z}\equiv\Es{z_0}=\Es{z_t}$ or $\covs(z,\chi)\equiv
\covs(z_0,\chi_0)=\covs(z_t,\chi_t)$ and equally for $\vars(z)$ and
$\Es{\nabla \chi\cdot\f D\nabla \chi}$.

\subsection{Results for detailed balance}
When the drift obeys detailed balance, i.e., $a(x)=D\nabla[\ln\ps(x)]$ for the steady-state density $\ps(x)$, %{changed here (do not use $\phi(x)$ anymore)}
% $D^{-1}a(x)=-\nabla\phi(x)$ and $\ps(x)={\rm e}^{-\phi(x)}$ {for a scalar field potential $\phi(x)$---actually do not need $\phi$ and could simply write $a(x)=D\nabla[\ln\ps(x)]$ as we do later},
the time-integrated steady-state covariance of $z_t$ is
bounded by \cite{Dechant2023PRL}
\begin{align}
\int_0^\infty\rmd t\,\covs(z_t,z_0) \, D^\chi_{\rm s} &\ge
\covs(z,\chi)^2\,,
\label{Dechant DB} 
\end{align}
where, using the notation $\Delta \chi_t\equiv\chi_{t+\Delta t}-\chi_t$,
\begin{align}
D^\chi_{\rm s} &\equiv\Es{\nabla \chi\cdot\f D\nabla \chi}=\lim_{\Delta t\to 0}\frac{\vars(\Delta \chi)}{2\Delta t}\,,\label{def Dchi}
\end{align}
and where for each $z$ there should be a particular $\chi$ (not
specified in \cite{Dechant2023PRL}) that
saturates the inequality. In Ref.~\cite{Dechant2023PRL} the authors
further define the ``correlation time'' as
\begin{align}
\tau^z =\frac{\int_0^\infty\rmd t\,\covs(z_t,z_0)}{\vars(z)}\,.
\end{align}
Since the time-integrated covariance looks a bit different in the
generalizations proved below and because such a notion of a correlation time only makes
sense for a single exponential decay, we refrain from interpreting $\tau^z$ as
a correlation time. 
However, while $\tau^z$ is some measure of the decay of
correlations, the bounds from
Ref.~\cite{Dechant2023PRL} and the generalizations below can more generally be
understood as inequalities concerning the variance of local
times/time-integrated densities. Since these variances are still fundamentally linked to the correlations in the system, we nevertheless consider the generalizations as ``thermodynamic correlation inequalities''.
% {\\Following sentence does not fit here? also eq 2 is always in equilibrium so do not need to mention it here---}We show below in Sec.~\ref{sec:spectral}, that in equilibrium we can obtain Eq.~\eqref{Dechant
%   DB} directly from a spectral-theoretic approach.

\subsection{Results for non-equilibrium steady states}
In NESS, the drift term $a(x)$ in Eq.~\eqref{SDE} can be decomposed into \cite{Seifert2012RPP,Dieball2022PRR}
\begin{align}
    a(x) = D\nabla\ln[\ps(x)] + \nus(x)\,,
\end{align}
where $\nus\colon\Omega\to\R^d$ is the local mean velocity that is zero in DB. Given $\ps$ and Eq.~\eqref{SDE}, we can express $\nus$ as $\nus(x)=a(x)-D\nabla\ln[\ps(x)]$. The steady-state (probability) current $\js\colon\Omega\to\R^d$ can be defined as $\js(x)\equiv\ps(x)\nus(x)$.

To generalize the results in Eq.~\eqref{Dechant DB} to NESS we introduce the notations 
\begin{align}
    \Sigma(\x) &\equiv \bnus(\x)\cdot\f D^{-1}\bnus(\x)\nonumber\\
    \sigmaE{\chi} &\equiv \int\rmd\x\,\chi(\x) \ps(\x)\frac{\Sigma(\x)}{\Es{\Sigma}}\nonumber\\
    \sigmavar(\chi) &= \sigmaE{\chi^2}-\sigmaE{\chi}^2\,.\label{defintion varSigma NESS}
\end{align}
Note that $\ps(\x)\Sigma(\x)/\Es{\Sigma}\ge 0$ and $\sigmaE{1}=1$ such that the construction $\sigmaE{\cdot}$ has the properties of a genuine expectation value. 
It was shown that \cite{Dechant2023PRL}
\begin{align}
\int_0^\infty\rmd t\,\covs(z_t,z_0)\left[D^\chi_{\rm s} + \sigmavar(\chi)\right]&\ge \covs(z,\chi)^2
\label{Dechant NESS}\,.
\end{align}
The term $\sigmavar(\chi)$ is not \textit{operationally accessible}
(i.e., not accessible from trajectories without first inferring $\ps$,
$a$ and $D$) but can be bounded by multiples of $\Es{\Sigma}$, e.g.,
by choosing a bounded observable $\chi$ \cite{Dechant2023PRL}. This,
in turn, allows for inferring bounds on $\Es{\Sigma}$ directly from
trajectories, which is physically meaningful since $\dot
S=\Es{\Sigma}$ is the entropy production rate in the system
\cite{Seifert2005PRL,Seifert2012RPP}. In this way, one can bound
the thermodynamic entropy production by the decay of correlations encoded in the
integral over the covariance in Eq.~\eqref{Dechant NESS}.  

\section{Summary of new results}

In this work, we generalize the thermodynamic correlation inequality in steady
states to finite times. Consider real or complex scalar observables $z_t=z(\x_t)$ and $y_t=y(\x_t)$ (square integrable functions $z,y:\Omega\to\C$). For complex observables, we  introduce complex conjugation in $\cov(a,b)\equiv\E{a^*b}-\E{a^*}\E{b}$ and equivalently in $\var$ and $D^y$. We denote a time
average up to time $t$ by $\overline{\cdots}_t$, i.e., $\overline{y}_t\equiv\int_0^ty_\tau\rmd\tau$. We have in DB
\begin{align*}
  D^y_{\rm s}\frac{t\,\vars(\overline{z}_t)}2&\ge \abs{\covs(y,z)-
\covs(\overline{y}_t,z_0)}^2\,,
\end{align*}
as well as in NESS
\begin{align*}
[D^y_{\rm s}+\Es{\Sigma(\x)}\var_\Sigma(y)]\frac{t\,\vars(\overline{z}_t)}2&\ge \abs{\covs(y,z)-\covs(\overline{y}_t,z_0)}^2\,.
\end{align*}

Moreover, we prove a thermodynamic correlation bound for transient dynamics (i.e., dynamics starting from any initial condition) relaxing to DB or NESS. With direct generalizations of the notations to time-dependent dynamics (details below), we have
\begin{align*}
&\left(D^y(t)+\overline{\E{\Sigma(\x_\tau,\tau)}\pvar_{\Sigma}^\tau(y)}_t-\frac1t[\var(y_t)-\var(y_0)]\right)\frac{t\var(\overline{z}_t)}2\nonumber\\
  &\qquad \ge\abs{\cov(\overline{y}_t,z_t)-\overline{\cov(y_\tau,z_\tau)}_t
    }^2,
\end{align*}
where $\pvar_{\Sigma}^\tau(y)\equiv\sigmaE{\abs{y_\tau-\E{y_\tau}}^2}$ is a pseudo variance of $y$, where $\sigmaE{\cdots}$ is the expectation with respect to the $\Sigma$-weighted
transient marginal density
$p(\x,t)\Sigma(\x,t)/\E{\Sigma(\x,t)}$. The term pseudo variance simply reflects that this is not a true ($\Sigma$-weighted) variance since generally $\E{y_\tau}\ne\sigmaE{y_\tau}$
Below we prove these
inequalities and discuss their saturation. We conclude with an outlook
of possible extensions and open problems.

\section{Spectral derivation for equilibrium and large time with
  insight into
 saturation}\label{sec:spectral}
Before we generalize the results, we first re-derive the result in
Eq.~\eqref{Dechant DB} following a spectral-theoretic approach (for a
spectral approach to fluctuations see, e.g.,
Ref.~\cite{Lapolla2020PRR}). This will give direct insight into the
saturation of the DB bound \eqref{Dechant DB} and show the emergence
of the mixture of short and long time scales emphasized in
Ref.~\cite{Dechant2023PRL}. For generality, we allow for
complex-valued observables, hence
$\cov(a,b)\equiv\E{a^*b}-\E{a^*}\E{b}$ for complex $a,b$ (note that $\cov(a,b)$
is not necessarily real; $\cov(a,a)=\var(a,a)$ is real, so are autocorrelations $\cov(a_{t_1},a_{t_2})$
under detailed balance).

In DB, an orthonormal basis of eigenfunctions exists (see, e.g., Ref.~\cite{Lapolla2020PRR}). Define
$\psi_i(\x)$, $i=0,1,2,\dots$ as the real, $\ps$-orthonormal
eigenfunction (i.e., $\int_\Omega \psi_i(x)\psi_k(x)\ps(x)dx=\delta_{ik}$) of the transposed operator $-L_x^T$, i.e., $-L_x^T\psi_i=\lambda_i\psi_i$ with eigenvalues $0=\lambda_0<\lambda_1\le\dots$ and $\psi_0=1$. Here, $L_x=-\nabla\cdot
D(\nabla-\{\nabla[\ln\ps(x)]\})$ and curly brackets denote that the operator only acts within the brackets. The eigenfunctions of $L_x$ are given by $\ps\psi_i$, i.e., $L_x\ps\psi_i=-\lambda_i\ps\psi_i$ \cite{Gardiner1985,Lapolla2020PRR}. Then we can write the
observables $y_\tau\equiv y(\x_\tau)$ and $z_\tau\equiv z(\x_\tau)$ with coefficients $c_i^{y,z}\in\C$ in the basis $\{\psi_i\}_{i\in\N_0}$ as %[the $\psi_i$ are real
  %(sums in covariance do not include $i=0$ as the mean was already subtracted)
\begin{align}
y_\tau &= \sum_{i\ge 0}c_i^y\psi_i(\x_\tau)\,,\qquad z_\tau = \sum_{i\ge
  0}c_i^z\psi_i(\x_\tau)\,,\end{align}
such that
\begin{align}
&\hspace{-4mm}\int_0^\infty\covs(z_{t'},z_0)\rmd
  t'=\sum_{i>0}\abs{c_i^z}^2\int_0^\infty\rmd
  t'\exp(-\lambda_it')=\sum_{i>0}\abs{c_i^z}^2/\lambda_i\,\in\R\,,
\end{align}
as well as
  \begin{align}
D^y_{\rm s}&\equiv\int\rmd\x\ps(\x)\{\nabla y(\x)\}^\dag\f D\nabla
y(\x)=\int\rmd\x y^*(\x)\nabla^T\ps(\x)\f D\nabla y(\x)\nonumber\\&=\int\rmd\x
y^*(\x)\hat
L_\x\ps(\x)y(\x)=\sum_{i>0}\abs{c_i^y}^2\lambda_i\in\R\,.
  \end{align}
We now apply the Cauchy-Schwarz inequality 
 \begin{align} 
&\Es{\abs{\sum_{i>0}\frac{c_i^z\psi_i(\x)}{\sqrt{\lambda_i}}}^2}\Es{\abs{\sum_{k>0}c_k^y\sqrt{\lambda_k}\psi_k(\x)}^2}\ge
   \abs{\Es{\left(\sum_{i>0}\frac{c_i^z\psi_i(\x)}{\sqrt{\lambda_i}}\right)^\dag\left(\sum_{k>0}c_k^y\sqrt{\lambda_k}\psi_k(\x)\right)}}^2,
 \end{align}
leading to (recall that $\Es{\cdots}=\int_\Omega\ps(x)\cdots dx$ and the $\psi_i$
are $\ps$-orthonormal)
 \begin{align}   
\int_0^\infty\covs(z_{t'},z_0)\rmd t'\,D^y_{\rm s}\ge \abs{\covs(y,z)}^2
%=\abs{\covs(z,y)}^2=\covs(z,y)\covs(y,z)
\,.\label{spectral proof easiest} 
\end{align} 
This proves Eq.~\eqref{Dechant DB} and concurrently also extends the
validity to complex observables and reversible Markov-jump dynamics (since the same spectral properties apply, see, e.g., Ref.~\cite{Lapolla2020PRR}). 
Moreover, we see that the connection between short and long times
corresponds to $1/\lambda_i$ and $\lambda_i$, and saturation for
\emph{any} given $z$ is achieved by choosing $y$ such that for all
$i>0$ we have $c_i^y=c_i^z/\lambda_i$ (which renders the predicted
equality of the supremum in \cite{Dechant2023PRL} exact).  We further
see immediately that $\tau^z$ is maximized for $y=z\propto\psi_1$,
where $\tau^z$ approaches the maximum $1/\lambda_1$. 
If we choose $z(x)$ to be any eigenfunction of $L_x^T$, the
inequality is saturated for $y=z$.

Due to the above insight that eigenfunctions provide saturation in DB, we
will henceforth generally allow for complex observables. Even though
the proof and insights from this section do \emph{not} generalize to
NESS, the results give a hint that complex observables may be
beneficial also for NESS, where eigenfunctions may become
complex (in general, $L_x$ can even become non-diagonalizable).
% (For NESS this proof does not work anymore [would also need to get $\Sigma(\x)$ somehow for Eq.~\eqref{Dechant NESS}].)

\section{From time-integrated correlation to variance of time-averaged density}
When generalizing the inequalities to finite times, we will find that
$\int_0^\infty\covs(z_{t'},z_0)\rmd t'$ does \emph{not} exactly reduce
to $\int_0^t\cov(z_{t'},z_0)\rmd t'$, but instead to $\int_0^t\rmd
t'(1-t'/t)\covs(z_{t'},z_0)$, since only the latter corresponds to a
variance which is accessible via the inequality recipe from
Ref.~\cite{Dieball2023PRL}. 

To see this, define for $z_\tau=z(\x_\tau)$ the time-averaged density $\overline{z}_t\equiv\frac{1}{t}\int_0^t z_\tau\rmd\tau$ and, using that $\covs(z_{t_1},z_{t_2})$ only depends on the time difference $t'=\abs{t_1-t_2}$, we obtain
\begin{align}
%\overline{z}_t&\equiv\frac{1}{t}\int_0^t z_\tau\rmd\tau\nonumber\\
t\,\vars(\overline{z}_t)&=\frac1t\int_0^t\rmd t_1\int_0^t\rmd
t_2\covs(z_{t_1},z_{t_2})\nonumber\\
&=2\int_0^t\rmd t'\left(1-\frac{t'}{t}\right)\covs(z_{t'},z_0)
%\overset{t\to\infty}\longrightarrow2\int_0^\infty\covs(z_{t'},z_0)\rmd t'
\nonumber\\
\Rightarrow \int_0^\infty\covs(z_{t'},z_0)\rmd t'&=\frac12\lim_{t\to\infty}t\,\vars(\overline{z}_t)
\,.
\end{align}
For complex observables $z\in\C$ we modify the above as follows [note
  that $\covs(z_{t'},z_0)$ is \emph{not} necessarily real in driven systems]
\begin{align}
t\,\vars(\overline{z}_t)
&=\frac1t\int_0^t\rmd t_1\int_0^t\rmd t_2\covs(z_{t_1},z_{t_2})\nonumber\\
&=\frac1{2t}\int_0^t\rmd t_1\int_0^t\rmd t_2\left[\covs(z_{t_1},z_{t_2})+\covs(z_{t_2},z_{t_1})\right]\nonumber\\
&=\frac1t\int_0^t\rmd t_1\int_0^t\rmd t_2\Re[\covs(z_{t_1},z_{t_2})]\nonumber\\
&=2\int_0^t\rmd t'\left(1-\frac{t'}{t}\right)\Re[\covs(z_{t'},z_0)]\,.\label{correlation variance complex} 
\end{align}

\section{Generalization of Eq.~\eqref{Dechant DB} to finite times}
\subsection{Proof}
We now generalize the result in Eq.~\eqref{Dechant DB} to finite
times, i.e., for now we still assume equilibrium dynamics starting
from the steady-state density $\ps$, but now consider a finite
trajectory length giving access to, e.g., $\int_0^t\rmd
t'\covs(\cdot,\cdot)$ instead of $\int_0^\infty\rmd
t'\covs(\cdot,\cdot)$. We will later generalize to dynamics
approaching both, DB or NESS, from general initial
conditions. Technically, the results of this section are contained in the later, more general, section. However, we first perform this simplest generalization to illustrate and explain our strategy for the proof.

We carry out the proof via the Cauchy-Schwarz inequality using
stochastic calculus, exactly as in Ref.~\cite{Dieball2023PRL}, but with a different choice of the integrals $A_t,B_t$. Consider the
It\^o equation \eqref{SDE} with $a(x)=D\nabla\ln[\ps(x)]$ and
define %evaluate
the following integrals for $y,z\in\C$
%blueC{now allow $y,z\in\C$; symmetry $t_1\leftrightarrow t_2$ should ensure that we obtain real parts as in Eq.~\eqref{correlation variance complex}---note that we do not multiplicative noise (but probably could..)}
\begin{align}
%\rmd\x_\tau&=\f a(\x_\tau)\rmd\tau+\bsig\rmd\f W_\tau\nonumber\\
%\f a(\x)&=\f D\nabla_\x\ln\ps(\x),\qquad \f D\equiv\bsig\bsig^T/2\nonumber\\
A_t&\equiv\frac1{\sqrt t}\int_0^t\{\nabla y_\tau\}\cdot\sigma\rmd \f
W_\tau\nonumber\\
B_t&\equiv\frac1{\sqrt t}\int_0^t(z_\tau-\Es{z})\rmd\tau=\sqrt t(\overline{z}_t-\Es{\overline{z}_t})\,,
\end{align}
such that
\begin{align}
\Es{\abs{A_t}^2}&=\frac1t\int_0^t\Es{\{\nabla y_\tau\}^\dag 2\f D\{\nabla y\}}\rmd\tau=\Es{\{\nabla y\}^\dag 2\f D\{\nabla y\}}=2D^y_{\rm s}\,,\nonumber\\
\Es{\abs{B_t}^2}&=t\,\vars(\overline{z}_t)
% \overset{t\to\infty}\longrightarrow 2\int_0^\infty\Re[\covs(z_{t'},z_0)]\rmd t', 
\,,\nonumber\\
\Es{A_t^*B_t}&=\frac{1}{t}\int_0^t\rmd\tau'\int_0^{\tau'}\Es{\{\nabla y_\tau\}^\dag\bsig\rmd \f W_\tau(z_{\tau'}-\Es{z})}\,.\label{setup DB}
\end{align}
Now, we use the 2-point Lemma from  Ref.~\cite{Dieball2022JPA} in the
form as in Eq.~(S37) in Ref.~\cite{Dieball2023PRL} (see Supplemental
Material therein), which
reads
%(should equally hold for complex $\f g, \mathcal U$)
\begin{align}
&\E{\f g(\x_\tau,\tau)\cdot\bsig(\x_\tau)\rmd\f W_\tau \mathcal U(\x_{\tau'},\tau')}\nonumber\\
&\qquad\qquad=\mathbbm{1}_{\tau<\tau'}\rmd\tau\int\rmd\x\int\rmd\x' \mathcal U(\x',\tau')\f g(\x,\tau)2\f D(\x)P(\x,\tau)\cdot\nabla_\x P(\x',\tau'|\x,\tau)\,.\label{Lemma_from_TUR}
\end{align}
Then the last line in Eq.~\eqref{setup DB} becomes (here, for $\ps$
initial conditions, notation $z_\Delta(x')=z(x')-\Es{z}$, and using $L_x\ps(\x)=\nabla_\x^T\ps(\x)\f D\nabla_x$)
\begin{align}
\Es{A_t^*B_t}&=\frac{1}{t}\int_0^t\rmd\tau'\int_0^{\tau'}\Es{\{\nabla y^*_\tau\}\cdot\bsig\rmd \f W_\tau(z_{\tau'}-\Es{z})}\nonumber\\
&=\frac1{t}\int_0^t\rmd\tau'\int_0^{\tau'}\rmd\tau\int\rmd\x\int\rmd\x'\{\nabla_\x y^*(\x)\}^Tz_\Delta(\x')2\f D\ps(\x)\nabla_\x G(\x',\tau'-\tau|\x)\nonumber\\
&=2\int\rmd\x\int\rmd\x'\int_0^t\rmd t'\left(1-\frac{t'}t\right)\{\nabla_\x y^*(\x)\}^Tz_\Delta(\x')\f D\ps(\x)\nabla_\x G(\x',t'|\x)
\nonumber\\
&=-2\int\rmd\x\int\rmd\x'\int_0^t\rmd t'\left(1-\frac{t'}t\right)y^*_\Delta(\x)z_\Delta(\x')L_x\ps(\x)G(\x',t'|\x)\nonumber\\
&\overset{\rm DB}=-2\int\rmd\x\int\rmd\x'\int_0^t\rmd t'\left(1-\frac{t'}t\right)y^*_\Delta(\x)z_\Delta(\x')L_x G(\x,t'|\x')\ps(\x')\nonumber\\
&\overset{\rm FPE}=-2\int\rmd\x\int\rmd\x'y^*_\Delta(\x)z_\Delta(\x')\ps(\x')\int_0^t\rmd t'\left(1-\frac{t'}t\right) \partial_{t'}G(\x,t'|\x')\nonumber\\
&=-2\int\rmd\x\int\rmd\x'y^*_\Delta(\x)z_\Delta(\x')\ps(\x')\times\nonumber\\
&\qquad\bigg[\underbrace{\left(1-\frac{t}t\right)}_{=0}G(\x,t|\x')-\underbrace{G(\x,0|\x')}_{\delta(\x-\x')}-\int_0^t\rmd t'G(\x,t'|\x')\underbrace{\partial_{t'}\left(1-\frac{t'}t\right)}_{-1/t}\bigg]\nonumber\\
&=2\left[\covs(y_0,z_0)-\frac1t\int_0^t\covs(y_{t'},z_0)\rmd t'\right]\,,\label{detailed balance calculation}
\end{align}
where $G(\x,t'|\x')$ is the transition probability density of $x_t$, DB denotes that we exploited detailed balance
$G(\x',t'|\x)\ps(\x)=G(\x,t'|\x')\ps(\x')$, and FPE that we inserted the Fokker-Planck equation $L_x G(\x,t'|\x')=\partial_{t'}G(\x,t'|\x')$ \cite{Gardiner1985,Risken1989}. 

Applying the Cauchy-Schwarz inequality to Eq.~\eqref{detailed balance
  calculation} finally gives
\begin{align}
\Es{A_t^2}\Es{B_t^2}&\ge \Es{A_t^*B_t}\Es{A_tB^*_t}\nonumber\\
\Rightarrow\quad D^y_{\rm s}\frac{t\,\vars(\overline{z}_t)}2&\ge \abs{\covs(y,z)-\frac1t\int_0^t\covs(y_{t'},z_0)\rmd t'}^2\,.\label{result finite time DB}
\end{align}
Note that we can write the last term equivalently as 
$\int_0^t\covs(y_{t'},z_t)\rmd t'=\int_0^t\covs(y_{t'},z_0)\rmd t'$ or as
$\frac1t\int_0^t\covs(y_{t'},z_0)\rmd t'=\covs(\overline{y}_t,z_t)$
(this will be useful for comparing to the generalizations below).

Revisiting the result from Ref.~\cite{Dechant2023PRL} for the
finite-time result we see that $D^y_{\rm s}$ is still the same as in Eq.~\eqref{Dechant DB}, but the
other term on the left-hand side is no longer a ``correlation time''
but instead the variance of a time-integrated density, and the
right-hand side now also displays some dependence on unequal-time
correlations. 

\subsection{Saturation conditions}
Consider $y_\tau=z_\tau=\psi_k(\x_\tau)$ for some $k>0$ (we choose $\psi_k(\x_\tau)$
to be $L_2(\ps)$-normalized s.t.\ $\vars(\psi_k)=1$; note that so far $\lambda_k\in\R$, hence $\cov(z_{t'},z_0)\in\R$). Using Eq.~\eqref{spectral proof easiest} as
well as Ref.~\cite[Eq.~(B4)]{Lapolla2020PRR}, we get $D^y_{\rm s}=\sum_{i>0}\abs{c_i^y}^2\lambda_i=\lambda_k$ and
\begin{align}
\frac{t\,\vars(\overline{z}_t)}2&=\int_0^t\rmd t'\left(1-\frac{t'}t\right)\cov(z_{t'},z_0)\nonumber\\
&=\int_0^t\rmd
t'\left(1-\frac{t'}t\right)\rme^{-\lambda_kt'}\overset{\footnotesize\text{
    Ref.~\cite{Lapolla2020PRR}}}=\frac1{\lambda_k}\left(1-\frac{1-\rme^{-\lambda_kt}}{\lambda_kt}\right)\nonumber\\
\covs(y,z)-\frac1t\int_0^t\covs(y_{t'},z_0)\rmd
t'&=1-\frac1t\int_0^t\rme^{-\lambda_kt'}\rmd
t'=1-\frac{1-\rme^{-\lambda_kt}}{\lambda_kt},
\label{finite time DB  spectral}
\end{align}
and by Eq.~\eqref{result finite time DB} we have 
\begin{align}
\lambda_k\frac1{\lambda_k}
\left(1-\frac{1-\rme^{-\lambda_kt}}{\lambda_kt}\right) & \ge \left(1-\frac{1-\rme^{-\lambda_kt}}{\lambda_kt}\right)^2\nonumber\\
\implies 1-\frac{1-\rme^{-\lambda_kt}}{\lambda_kt}&< 1\quad \forall
t\in (0,\infty)\,.\label{no saturation}
\end{align}
Therefore, we obtain equality only for $t\to\infty$ when $y=z=\psi_k$
with $k>0$. Complete saturation for finite times is probably not
feasible since this requires $A_t$ and $B_t$ in Eq.~\eqref{setup DB}
to be completely correlated (see saturation for Cauchy-Schwarz step of
the proof as in Ref.~\cite{Dieball2023PRL}). There is no (at least no
obvious) choice to achieve this, since $A_t$ is a $\rmd\f
W_\tau$-integral while $B_t$ is a (stochastic) $\rmd\tau$-integral.

\subsection{Convergence to Eq.~\eqref{Dechant DB}}\label{sec:convergence to long time}
For the special case of $y=z=\psi_k$, we confirm convergence towards
the equilibrium result in Eq.~\eqref{finite time DB spectral}, also
considering the two sides of the inequality individually. In
particular, we see that the LHS approaches the $t\to\infty$ result as
$1-\frac{1-\rme^{-\lambda_kt}}{\lambda_kt}\to 1$, while the RHS does
so with $(1-\frac{1-\rme^{-\lambda_kt}}{\lambda_kt})^2\to 1$.
%Maybe compare this with transients later (think of the result in \cite{Dieball2022JPA} where we saw that often it takes longer to forget transients than it does to approach LD variance from $\ps$ IC).

%\section{Generalization of Eq.~\eqref{Dechant NESS} to finite times
%  and of Eqs.~\eqref{Dechant DB} and \eqref{Dechant NESS} to transient
%  dynamics}
\section{Generalization to finite times, irreversible drift, and transient dynamics}

\subsection{Setup}
We now do the derivation for the most general version (transients
towards NESS) and later discuss the three different cases (transients
to DB, finite-time NESS, transients to NESS). We have the
Fokker-Planck operator $L_\x$ and its dual-reverse operator
$L_\x^{-\js}$ \cite{Dieball2022PRL,Dieball2022PRR} such that, 
\begin{align}
L_\x\ps(\x)&=\nabla_\x\cdot\f D\ps(\x)\nabla_\x-\js(\x)\cdot\nabla_\x\nonumber\\
L_\x^{-\js}\ps(\x)&\equiv\nabla_\x\cdot \f D\ps(\x)\nabla_\x+\js(\x)\cdot\nabla_\x,
\label{FPE}  
\end{align}
with transition probability densities $G(\x',t'|\x)$ and
$G^{-\js}(\x,t'|\x')$, respectively, that obey dual reversal symmetry
$G(\x',t'|\x)\ps(\x)=G^{-\js}(\x,t'|\x')\ps(\x')$ \cite{Dieball2022PRL,Dieball2022PRR}. We denote the
marginal density at $t$ for a general initial condition $p_0(x_0)$ as
$p(\x,t)\equiv\int_\Omega G(\x,t|x_0)p_0(x_0)\rmd\x_0$ with probability
current $j(\x,t)=[-D\ps(\x)\nabla_\x\ps(x)^{-1}+\nus(\x)]p(\x,t)$
and expectation $\langle \cdots\rangle\equiv \int_\Omega \cdots
p(\x,t)\rmd x$. We further introduce the 
local mean velocity $\bnu(\x,t)\equiv j(\x,t)/p(\x,t)$ as well as
$\Sigma(\x,\tau)\equiv\bnu(\x,\tau)\cdot \f D^{-1}\bnu(\x,\tau)$ such
that $\E{\Sigma(\x_\tau,\tau)}\ge 0$ is the total entropy production
rate \cite{Seifert2012RPP}. We further introduce the $\Sigma$-weighted expectation
\begin{align}
\E{\cdots}_\Sigma\equiv\frac{\int_\Omega\cdots
  p(\x,t)\Sigma(\x,t)\rmd\x}{\int_\Omega p(\x,t)\Sigma(\x,t)\rmd\x}=\frac{\int_\Omega\cdots
  p(\x,t)\Sigma(\x,t)\rmd\x}{\E{\Sigma(\x,t)}},
  \label{wexp}
\end{align}
with corresponding ``pseudo variance''
$\pvar_{\Sigma,C}^\tau(\cdots)\equiv\E{|\cdots-C_\tau|^2}_\Sigma$ (only a
true variance for $C_\tau=\E{\cdots}_\Sigma$).

\subsection{Proof}
% Consider Eq.~\eqref{setup DB} (now allow $\bnu\ne\f 0$) with
% check everywhere below that we use 'pvar' and 'var' correctly\\}
Generalize the definition in Eq.~\eqref{setup DB} for $\nu\ne0$ to
\begin{align}
A_t&\equiv\frac1{\sqrt t}\int_0^t\left[\{\nabla y_\tau\}-\f D^{-1}\bnu(\x_\tau,\tau)(y_\tau-C_\tau)\right]\cdot\bsig\rmd \f W_\tau,
\end{align}  
such that
\begin{align}
\frac{\E{\abs{A_t}^2}}{2}&=D^y(t)+\frac1t\int_0^t\rmd\tau\E{\Sigma(\x_\tau,\tau)}\pvar_{\Sigma,C}^\tau(y_\tau)\nonumber\\&-\frac1t\int_0^t\rmd\tau\Elr{\bnu(\x_\tau,\tau)\left[(y_\tau-C_\tau)^*\cdot\nabla y_\tau+(y_\tau-C_\tau)\cdot\nabla y_\tau^*\right]}\\
&=D^y(t)+\frac1t\int_0^t\rmd\tau\E{\Sigma(\x_\tau,\tau)}\pvar_{\Sigma,C}^\tau(y_\tau)-\frac1t\int_0^t\rmd\tau\Elr{\bnu(\x_\tau,\tau)\cdot\nabla\abs{y_\tau-C_\tau}^2},\nonumber
\end{align}
where $D^y(t)\equiv\E{\{\nabla y_t^*\}\cdot D\{\nabla y_t\}}$, and
\begin{align}
\int_0^t\rmd\tau & \Elr{\bnu(\x_\tau,\tau)\cdot\nabla\abs{y_\tau-C_\tau}^2}=-\int_0^t\rmd\tau\int\rmd\x\abs{y(\x)-C_\tau}^2\nabla_\x\cdot\f j(\x,\tau)\nonumber\\
&=\int_0^t\rmd\tau\int\rmd\x\abs{y(\x)-C_\tau}^2\partial_\tau p(\x,\tau)\,.
\label{setup transients 2}
\end{align}
%\begin{align}
%L_\x\ps(\x)&=\nabla_\x^T\f D\ps(\x)\nabla_\x-\js(\x)\cdot\nabla_\x\nonumber\\
%L_\x^{-\js}\ps(\x)&\equiv\nabla_\x^T\f D\ps(\x)\nabla_\x+\js(\x)\cdot\nabla_\x\nonumb%er\\
%G(\x',t'|\x)\ps(\x)&=G^{-\js}(\x,t'|\x')\ps(\x')\nonumber\\
%\f j(\x,\tau)&=\bj_\x P(\x,\tau)=\left[-\f D\ps(\x)\nabla_\x\ps^{-1}(\x)+\bnus(\x)\right]P(\%x,\tau)
%\nonumber\\
%\bnus(\x)&=\frac{\js(\x)}{\ps(\x)}\nonumber\\
%\bnu(\x,\tau)&=\frac{\f j(\x,\tau)}{p(\x,\tau)}\nonumber\\
%\f D\ps(\x)\nabla_\x\frac{P(\x,\tau)}{\ps(\x)}&=\bnus(\x)P(\x,\tau)-\f j(\x,\tau)%\nonumber\\
%\Sigma(\x,\tau)&\equiv\bnu(\x,\tau)\cdot \f D^{-1}\bnu(\x,\tau)\nonumber\\
%\E{\Sigma(\x_\tau,\tau)}\pvar_{\Sigma,C}^\tau(y_\tau)&\equiv\int\rmd\x p(\x,\tau)\Sigm%a(\x,\tau)\abs{y(\x)-C_\tau}^2\label{setup transients 1}
%\end{align}
Now for transients, where $\partial_\tau p(\x,\tau)\ne 0$, we will
have to choose $C_\tau=\E{y_\tau}$ to later perform the necessary simplifications. We will denote for this choice the pseudo variance $\pvar_{\Sigma,C}^\tau$ by omitting the $C$, i.e., $\pvar_\Sigma^\tau(y_\tau)\equiv\sigmaE{\abs{y_\tau-\E{y_\tau}}^2}$ (not a true variance since generally $\E{\cdots}\ne\sigmaE{\cdots}$).

For NESS we can choose $C_\tau$ more freely  and opt for
$C_\tau=C=\E{y}_\Sigma$ corresponding to the results in
Eq.~\eqref{Dechant NESS}, which is more beneficial for further bounds since it is a true variance 
(later this will allow for application of Popoviciu's inequality as in Ref.~\cite{Dechant2023PRL}). With this choice, denoted by $\var_\Sigma(y)\equiv\sigmaE{\abs{y-\E{y}_\Sigma}^2}$, we have for NESS
\begin{align}
% \E{\abs{A_t}^2}&=2D^y(t)+\frac2t\int_0^t\rmd\tau\E{\Sigma(\x_\tau,\tau)}\var_\Sigma^\tau(y_\tau),
\E{\abs{A_t}^2}&=2D_{\rm s}^y(t)+\frac2t\int_0^t\rmd\tau\Es{\Sigma(\x_\tau)}\var_\Sigma(y)=2D_{\rm s}^y(t)+\frac2t\Es{\Sigma(\x)}\var_\Sigma(y)\,.
\label{Asq NESS}
\end{align}
For transients we choose $C_\tau=\E{y_\tau}$ and note that 
\begin{align}
\int_0^t\rmd\tau & \int\rmd\x\abs{y(\x)-C_\tau}^2\partial_\tau p(\x,\tau)\nonumber\\
&=\int_0^t\rmd\tau\partial_\tau\int\rmd\x P(\x,\tau)(y_\tau^*y_\tau-C_\tau^*y_\tau-y_\tau^*C_\tau+C_\tau^*C_\tau)\nonumber\\
&=\int_0^t\rmd\tau\partial_\tau\left(\E{y_\tau^*y_\tau}-C_\tau^*\E{y_\tau}-\E{y_\tau^*}C_\tau+C_\tau^*C_\tau\right)\nonumber\\
&=\int_0^t\rmd\tau\partial_\tau \var(y_\tau)\nonumber\\
&=\var(y_t)-\var(y_0)\,,
\label{setup transients 3}
\end{align}
where the first equality follows from $\partial_\tau y(\x)=0$ and from
the choice $C_\tau=\E{y_\tau}$ we get
$\E{\partial_\tau\abs{y(\x_\tau)-C_\tau}^2}=-\E{y^*\dot C-y{\dot
    C}^*+C{\dot C}^*+C^*\dot C}=-\E{y^*}\dot C-\E{y}{\dot C}^*+C{\dot
  C}^*+C^*\dot C=0$, which illustrates why we opted for this choice
  for transients.
  % (although a constant $C_\tau=C$ would also allow this step).
Altogether we obtain for transients for $C_\tau=\E{y_\tau}$ that
\begin{align}
\E{\abs{A_t}^2}&=2D^y(t)+\frac2t\int_0^t\rmd\tau\E{\Sigma(\x_\tau,\tau)}\pvar_{\Sigma}^\tau(y_\tau)-\frac2t\left[\var(y_t)-\var(y_0)\right]\,.\label{Asq transients} 
\end{align}

We now evaluate $\E{A_t^*B_t}$ via Lemma~\eqref{Lemma_from_TUR} as
(note that $y_\Delta(\x,\tau)=y(\x)-\E{y_\tau}$ now depends on time)
\begin{align}
&\E{A_t^*B_t}=\frac{1}{t}\int_0^t\rmd\tau'\int_0^{\tau'}\Elr{\left[\{\nabla y_\tau\}-\f D^{-1}\bnu(\x_\tau,\tau)(y_\tau-\E{y_\tau})\right]^*\cdot\bsig\rmd \f W_\tau(z_{\tau'}-\E{z_{\tau'}})}\nonumber\\
&=\frac1{t}\int_0^t\rmd\tau'\int_0^{\tau'}\rmd\tau\int_\Omega\rmd\x\int_\Omega\rmd\x'\left[\{\nabla_x y(\x)\}-\f D^{-1}\bnu(\x,\tau)y_\Delta^*(\x,\tau)\right]^\dag
\times\nonumber\\&\qquad\qquad
z_\Delta(\x',\tau')2\f Dp(\x,\tau)\nabla_\x G(\x',\tau'-\tau|\x)\nonumber\\
&=-\frac2{t}\int_0^t\rmd\tau'\int_0^{\tau'}\rmd\tau\int_\Omega\rmd\x\int_\Omega\rmd\x'y_\Delta^*(\x,\tau)z_\Delta(\x',\tau')\left[\nabla_x+\f D^{-1}\bnu(\x,\tau)\right]^T
\cdot\nonumber\\&\qquad\qquad
\frac{p(\x,\tau)}{\ps(\x)}\f D \ps(\x)\nabla_\x G(\x',\tau'-\tau|\x)\nonumber\\
&=-\frac2{t}\int\rmd_{\tau,\x,y,z}\bigg[\underbrace{\f D \ps(\x)\left\{\nabla_\x\frac{p(\x,\tau)}{\ps(\x)}\right\}+p(\x,\tau)\bnu(\x,\tau)}_{=\bnus(\x)p(\x,\tau)}+\frac{p(\x,\tau)}{\ps(\x)}\nabla_\x^T\f D\ps(\x)\bigg]
\cdot\nonumber\\&\qquad\qquad
\nabla_\x G(\x',\tau'-\tau|\x)\nonumber\\
&=-\frac2{t}\int\rmd_{\tau,\x,y,z}\frac{p(\x,\tau)}{\ps(\x)}\underbrace{\left[\js(\x)+\nabla_\x^T \f D\ps(\x)\right]\cdot\nabla_\x G(\x',\tau'-\tau|\x)}_{=L_\x^{-\js}\ps(\x)G(\x',\tau'-\tau|\x)=L_\x^{-\js}\ps(\x')G^{-\js}(\x,\tau'-\tau|\x')}\nonumber\\
&=-\frac2{t}\int\rmd_{\tau,\x,y,z}\frac{p(\x,\tau)}{\ps(\x)}\partial_{\tau'} G^{-\js}(\x,\tau'-\tau|\x')\ps(\x')\nonumber\\
&=-\frac2{t}\int\rmd_{\tau,\x,y,z}\partial_{\tau'} G(\x',\tau'-\tau|\x)p(\x,\tau)\,,\label{calculation transients 1}
\end{align}
where in the third line we introduced the shorthand notation for the
multiple integrals
\begin{align}
\int\rmd_{\tau,\x,y,z}\cdots\equiv \int_0^t\rmd\tau'\int_0^{\tau'}\rmd\tau\int_\Omega\rmd\x\int_\Omega\rmd\x'y_\Delta^*(\x,\tau)z_\Delta(\x',\tau')\cdots\,.
\end{align}  
Note that terms
$\E{z_{\tau'}}\partial_{\tau'}\int_\Omega\rmd\x'G(\x',\tau'-\tau|\x)p(\x,\tau)=\E{z_{\tau'}}\partial_{\tau'}1=0$
do not contribute since $\E{z_{\tau'}}$ is independent of $\x'$ such
that we can use normalization. Therefore, we have no problems from the
time-dependence of $y_\Delta,z_\Delta$ and get (by performing an
  integration by parts with $\partial_{\tau'}\mathbbm{1}(\tau\le\tau')=\delta(\tau'-\tau)$ in the second equality)
\begin{align}
\E{A_t^*B_t}&=-\frac2{t}\int_0^t\rmd\tau'\int_0^t\rmd\tau\mathbbm{1}(\tau\le\tau')\partial_{\tau'}\cov(y_\tau,z_{\tau'})\nonumber\\
&=\frac2{t}\int_0^t\rmd\tau\left(\int_0^t\rmd\tau'\delta(\tau'-\tau)\cov(y_\tau,z_{\tau'})-\left[\mathbbm{1}(\tau\le\tau')\cov(y_\tau,z_{\tau'})\right]_{\tau'=0}^{\tau'=t}\right)\nonumber\\
&=\frac2{t}\int_0^t\rmd\tau\,\cov(y_\tau,z_\tau-z_t) 
\nonumber\\&
=-2\cov(\overline{y}_t,z_t)+\frac2{t}\int_0^t\rmd\tau\,\cov(y_\tau,z_\tau)\,.
\label{calculation transients 2}
\end{align}
Since $\E{\abs{B_t}^2}=t\var(\overline{z}_t)$ as in Eq.~\eqref{setup
  DB}, we are done with all calculations and now list the respective results.

\subsection{Result for NESS}
From Eqs.~\eqref{Asq NESS} and \eqref{calculation transients 2} we
obtain via Cauchy-Schwarz the correlation inequality
\begin{align}
[D^y_{\rm s}+\Es{\Sigma(\x)}\var_\Sigma(y)]\frac{t\,\vars(\overline{z}_t)}2&\ge \abs{\covs(y,z)-\covs(\overline{y}_t,z_t)}^2\,.\label{result finite time NESS}
\end{align}
Exactly as Eq.~\eqref{result finite time DB} approaches
Eq.~\eqref{Dechant DB} for $t\to\infty$, we see that Eq.~\eqref{result
  finite time NESS} approaches Eq.~\eqref{Dechant NESS}. Note that one
can bound $\var_\Sigma(y)$ \blue{without knowledge of the underlying degrees of freedom in $\x$}, e.g., by choosing bounded $y$. \blue{In this way, one obtains an operationally accessible (i.e., accessible without measuring all degrees of freedom in $\x$)} lower bound for $\Es{\Sigma(\x)}$ \cite{Dechant2023PRL} which is the steady-state entropy production rate \cite{Seifert2012RPP}. \blue{Explicitly, if one can bound $\var_\Sigma(y)\le C(y)$ in terms of some constant or function $C(y)$ (see Ref.~\cite{Dechant2023PRL} for the example of bounded $y$), the bound on the entropy production rate $\Es{\Sigma(\x)}$ reads
\begin{align}
    \Es{\Sigma(\x)}
    \ge
    % \frac1{\var_\Sigma(y)}
    \frac1{C(y)}
    \left[\frac{2\abs{\covs(y,z)-\covs(\overline{y}_t,z_t)}^2}{t\,\vars(\overline{z}_t)} - D^y_{\rm s}\right]
    \,.\label{bound EPR NESS}
\end{align}%}
%\noindent \blue{
To obtain a useful bound on $\Es{\Sigma(\x)}$, the right-hand side of Eq.~\eqref{bound EPR NESS} has to be positive, which means that the DB result in Eq.~\eqref{result finite time DB} has to be violated (which is possible in NESS as discussed for $t\to\infty$ in Ref.~\cite{Dechant2023PRL}). 

The results in Eqs.~\eqref{result finite time NESS} and \eqref{bound EPR NESS} may generally feature a complicated time dependence. However, since they are generalizations of Eq.~\eqref{result finite time DB}, we expect the inequalities to be sharper for long times, although we only see this explicitly for the special case investigated in Eq.~\eqref{no saturation}.

Contrasting the time-dependence of thermodynamic uncertainty relations \cite{Manikandan2020PRL,Dieball2023PRL}, the short-time limit $t\to 0$ does \emph{not} appear promising for correlation inequalities (again, see Eq.~\eqref{no saturation} for an instructive special case).}

\subsection{Result for transients (with or without driving)}
From Eqs.~\eqref{Asq transients} and \eqref{calculation transients 2}
we obtain via Cauchy-Schwarz with $C_\tau=\E{y_\tau}$ the bound
\begin{align}
&\left(D^y(t)+\frac1t\int_0^t\rmd\tau\E{\Sigma(\x_\tau,\tau)}\pvar_{\Sigma}^\tau(y)-\frac1t[\var(y_t)-\var(y_0)]\right)\frac{t\var(\overline{z}_t)}2\nonumber\\
  &\qquad \ge\abs{\cov(\overline{y}_t,z_t)-\overline{\cov(y_\tau,z_\tau)}_t
    }^2\,,\label{result transients}
\end{align}
where we introduced the time-averaged equal-time covariance
$\overline{\cov(y_\tau,z_\tau)}_t\equiv
\frac1{t}\int_0^t\rmd\tau\,\cov(y_\tau,z_\tau)$. 

Note that we can bound
% {added C here---think of adopting this in $\tau$ notation $\var_{\Sigma}^\tau(y)$ for transitents with $\tau$ and the true variance for NESS without $\tau$?---\textbf{consider only writing true variances as $\var$ to avoid confusion}} 
$\pvar_{\Sigma,C}^\tau(y)$ in terms of $\E{\Sigma(\x_\tau,\tau)}$, e.g., by choosing bounded $y$%
% (although this is then \emph{not} a proper variance anymore, s.t., the
% trivial bound is worse by a factor of $4$ {state the trivial bounds to explain this factor of 4})
, to obtain a lower bound
% for $\E{\Sigma(\x_\tau,\tau)}=\partial_\tau\Delta S_{\rm tot}$. 
for the (total) entropy production rate $\E{\Sigma(\x_\tau,\tau)}=\dot S(\tau)$. However, in the case of a pseudo variance, the bounds may be more loose than for a true variance. Concretely, for observables that map to a bounded interval, $y_\tau\in[c,c+\Delta]$, we have by Popoviciu's inequality for variances that $\var_\Sigma(y_\tau)\le\Delta^2/4$, while for a pseudo variance with $y_\tau,C_\tau\in[c,c+\Delta]$ we generally only have $\pvar_{\Sigma,C}^\tau\equiv\sigmaE{\abs{y_\tau-C_\tau}^2}\le\Delta^2$.

Letting $t\to\infty$ in Eq.~\eqref{result transients} we should approach the steady-state result
Eq.~\eqref{Dechant NESS}.
However, note that this convergence is not obvious since (unlike the $t\to\infty$-convergence of the finite-time result in  Sec.~\ref{sec:convergence to long time}) it does not only involve the relaxation time scales. Instead, the initial condition explicitly enters the time averages and is therefore not quickly forgotten (only slowly decays via the $1/t$ in front of the average; see also Ref.~\cite{Dieball2022JPA} for an example where forgetting the initial condition takes longer than approaching a $t\to\infty$-solution from a finite-time result). As an extreme case, consider an initial condition starting in a point, $p(x,t=0)=\delta(x-x_0)$, where $\nu(x,t)$ and hence $\Sigma(x,t)$ diverges as $x\to x_0$, $t\to0$ and hence the left-hand side of
Eq.~\eqref{result transients} diverges for all $t>0$, such that the long-time result in Eq.~\eqref{Dechant NESS} will seemingly never be approached.
% but it seems that we require even (much)
% larger $t$, since the initial condition is not quickly forgotten. As an extreme case consider starting at a point such that the
% entropy production is infinite and in turn the left-hand side of
% Eq.~\eqref{result transients} is infinite for all times
% (Eq.~\eqref{Dechant NESS} perhaps may still apply after some long time 
% if the integrated correlations diverge as well, but this remains unclear).

\section{Outlook}
Using stochastic calculus and spectral theory, we gave direct proofs of several thermodynamic correlation inequalities for general
scalar (incl.\ complex-valued) observables and discussed their
saturation. Our results provide saturation conditions for the
inequalities derived in \cite{Dechant2023PRL} and generalize them 
to finite observation times and general initial conditions (i.e.\ to transients).

Our generalization of the bounds to finite times paves the way for
realistic applications. Further work may address experimental
applications (e.g., for colloidal particle tracking data
\cite{Dieball2025SM,Martinez2017SM,Ciliberto2017PRX,blickle2012realization} or single-molecule spectroscopy \cite{Lipman2003S,Dieterich2015NP,Ye2018NL,Tancredi2024NJP,Vollmar2024JPAMT}) \blue{or active systems \cite{Bechinger2016RMP,Pietzonka2019PRX,RobinBebon2025PRX}}. It will be interesting to see the practical relevance of the correlation bounds compared to other recent inference strategies (as summarized, e.g., in Ref.~\cite{Dieball2025JCP}).
More conceptual work could analyze the bounds on the dissipation in light of a correlation-dissipation tradeoff, complementary to tradeoffs of dissipation with, e.g., precision \cite{Barato2015JSP,Barato2016PRX}, transport \cite{Leighton2022PRL,Cai_transport} or adaptation \cite{Jefe}.
Further limits of the presented results, such as short-time limits or saturation close to equilibrium, could be studied in more detail in the future, and extension to Markov jump dynamics could be attempted.

% \red{In the revision, I added the references \cite{Deguenther2024PNASU,Otsubo2022CP,Leighton2023PRL,Boffi2024PNASU,Fyodorov2025PRL,Shiraishi2018PRL,CSL_4,Ye2018NL,Tancredi2024NJP,Martinez2017SM,Ciliberto2017PRX,Lipman2003S,Dieterich2015NP,Bechinger2016RMP,Pietzonka2019PRX,RobinBebon2025PRX}}

\section*{Acknowledgments}
Financial support from the European Research Council (ERC) under the European Union’s Horizon Europe research and
         innovation program (grant agreement No 101086182 to AG) is
         gratefully acknowledged.

\section*{References}
\bibliography{correlations}
\end{document}